\documentclass[useAMS,usenatbib,referee]{mn2e}
\usepackage{amsmath}
\usepackage[dvips]{graphics}
\usepackage {epsfig}
\title[]{Perturbative signature of substructures in strong gravitational lenses.}
\author[C. Alard]{C. Alard \thanks{E-mail:alard@iap.fr} \\
Institut d'Astrophysique de Paris, 98bis boulevard Arago, 75014
Paris}
\begin{document}
\label{firstpage}
\maketitle
\begin{abstract}
In the perturbative approach, substructures in the lens can be reduced
to their effect on the two perturbative fields $f_1$ and $\frac{d f_0}{d \theta}$.
A simple generic model of elliptical lens with a substructure situated near the critical
 radius is investigated in details.  Analytical
expressions are derived for each perturbative field, and basic properties are analyzed. The power spectrum
of the fields is well approximated by a power-law, resulting in significant tails at high frequencies.
Another feature of the perturbation by a substructure is that the ratio of the power spectrum at order $n$
 of the 2 fields $R_n$ is nearly 1. The ratio $R_n \simeq 1$ is specific to substructures, for instance 
an higher order distortion ($n>2$) but with auto-similar isophotes will result in $R_n \propto \frac{1}{n^2}$.  
Finally, the problem of reconstructing the perturbative field is investigated. Local field model are implemented and
fitted to maximize image similarity in the source plane. The non-linear optimization is greatly facilitated, since
in the perturbative approach the circular source solution is always known. Examples of images 
distortions in the subcritical regime due to substructures are presented, and analyzed for different source shapes. Provided enough images
 and signal is available, the substructure field can be identified confidently. These results suggests that
the perturbative method is an efficient tool to estimate the contribution of substructures to the mass
distribution of lenses.
\end{abstract}
\begin{keywords}
gravitational lensing-strong lensing
\end{keywords}

\section{Introduction.}
Images formed in the strong gravitational regime are very sensitive to the local variations of the lens deflection field.
In a circular potential the caustic is reduced to a single point and the image of a source in a singular situation is a circle.
When ellipticity is introduced, the caustic system is a serie of connected lines with typical diamond shape aspect. This system
presents cusps and folds singularities as predicted by the theory of singularities. Ellipticity is the first morphological deviation
from circular symmetry, and is a standard ingredient of lens models. However, higher order deviations from circular symmetry are significant
in practice, and mostly related to substructures or interaction at close range between galaxies, as observed in merging effects.
Not all lenses are mergers, but in general substructures are expected in the mass distribution of lenses (see for instance Klypin {\it et al.}
1999, Moore  {\it et al.} 1999, Ghigna  {\it et al.} 2000). The presence of substructures has several effects, modifying the optical depth (Horesh  {\it et al.} 2005,
 Meneghetti  {\it et al.} 2007), or altering the image flux (Bartelmann {\it et al} 1995, Keeton {\it et al} 2003, Mao  {\it et al} 2004, Maccio  {\it et al} 2006)
By analyzing the image flux for very small sources, like distant quasars it should be possible in principle to evaluate the contribution
of substructure to the lens deflection field. However, in practise small sources like quasars may be quite sensitive to microlensing by stars
in the lens galaxy, which complicates the analysis. 
This paper will investigate the effect of substructures for much larger sources, typically galaxies. In particular, 
the modifications of the image morphology due to the perturbing field of the substructures will be investigated in details.
\section{Perturbative description of the effects of substructure on arcs.}
\begin{figure}
\centering{\epsfig{figure=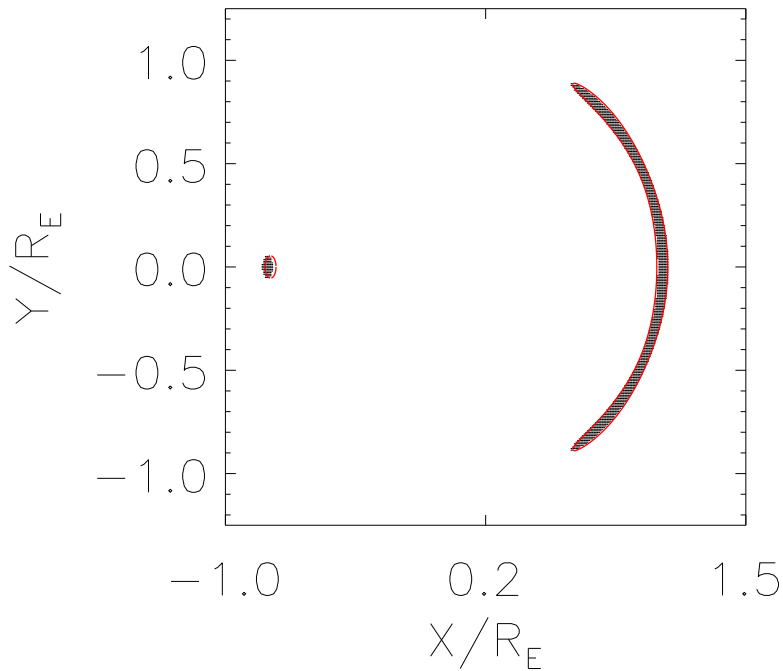,width=12cm}}
\caption{Near cusp configuration for an elliptical NFW profile without substructure. The source
 is circular with diameter $R_S=0.05 R_E$. The black images have been obtained by direct
 ray-tracing. The red lines are the contours of the circular
 source obtained using the perturbative formalism.}
\label{plot01}
\end{figure}

\begin{figure}[b]
\centering{\epsfig{figure=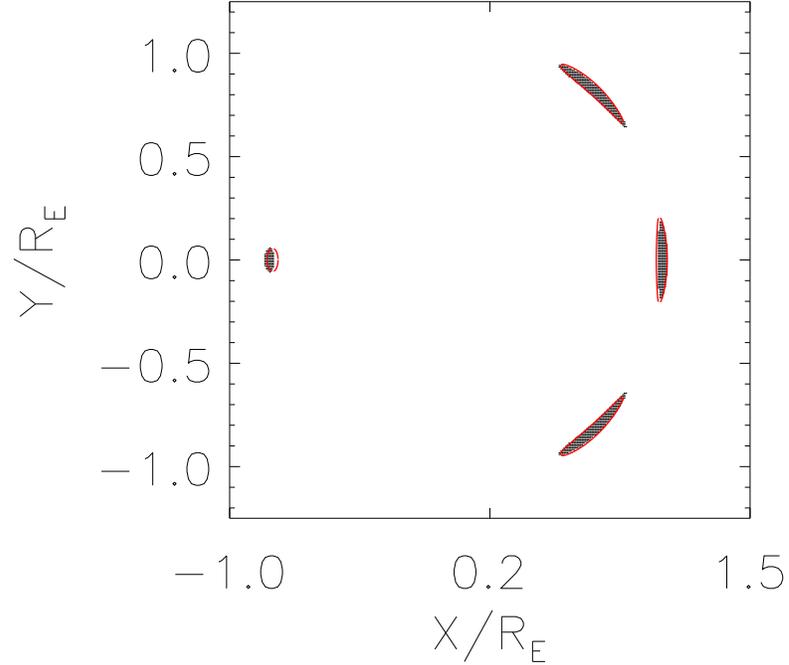,width=12cm}}
\caption{Same source position, but in addition to the elliptical deflector a substructure
with mass equal to one percent of the elliptical lens. The perturbator is located on the X-axis
at a distance $r_p=1.3 R_E$.}
\label{plot02}
\end{figure}

 Small substructures with mass of about a percent of the main deflector can
 have a major influence on  the shape of gravitational arcs. To be effective, the
 perturbator must be located near the critical line, and in a area of
 image formation ($|\frac{d f_0}{d \theta}| < {\rm source \ radius}$).
 The effect of substructure is illustrated in Fig. (~\ref{plot01}) and Fig. (~\ref{plot02}).
 In  Fig. (~\ref{plot01}) we are in the cusp caustic regime of an elliptical lens, while
 in Fig. (~\ref{plot02}) the perturbative field of a  sub-substructure is added to the elliptical
 lens. The field of the substructure changes dramatically the shape of the images, the arc
 is broken in 3 images, which is a situation typical of a sub-critical regime. 
  In both case the perturbative approximation is over-plotted on the image
 of the sources obtained using ray-tracing. To obtain the image contours by the 
perturbative method, the derivatives of the lensing
potential are estimated at $r=R_E$ (Einstein radius), providing the fields, $f_1$ and $\frac{d f_0}{d \theta}$
which are directly introduced in Eq. (12) (Alard 2007).
 As expected the perturbative method gives accurate results
 in the elliptical case, but also in the perturbed elliptical case, which suggests that the perturbative
 method is an efficient tool to study the perturbations of arcs by substructures. 
 More detailed investigations of lens with substructures using the perturbative 
 method are available in Peirani {\it et al.} 2008.
 In the perturbative approach all the information on the deflection potential is 
 in two fields, $f_1(\theta)$ and $\frac{d f_0}{d \theta}$. Thus it is sufficient to evaluate
 the effect of the substructure on these two fields, and this will be the main goal of the
 present paper.
\subsection{Field perturbation induced by substructures.}
\begin{figure}
\centering{\epsfig{figure=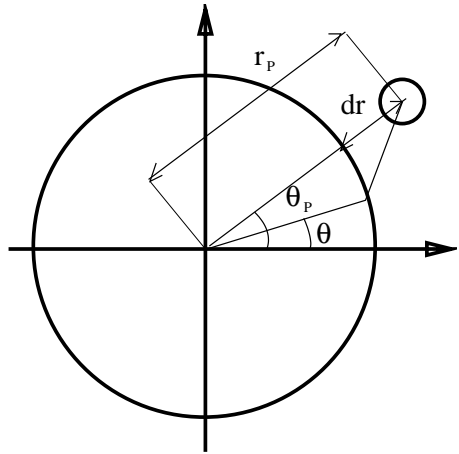,width=6cm}}
\caption{Geometry of the perturbation introduced by the substructure. The great circle
is the Einstein radius (normalized to unity) of the un-perturbed distribution, while the small circle represents
the substructure situated at $r_p=1+dr$.}
\label{plot1}
\end{figure}

\begin{figure}
\centering{\epsfig{figure=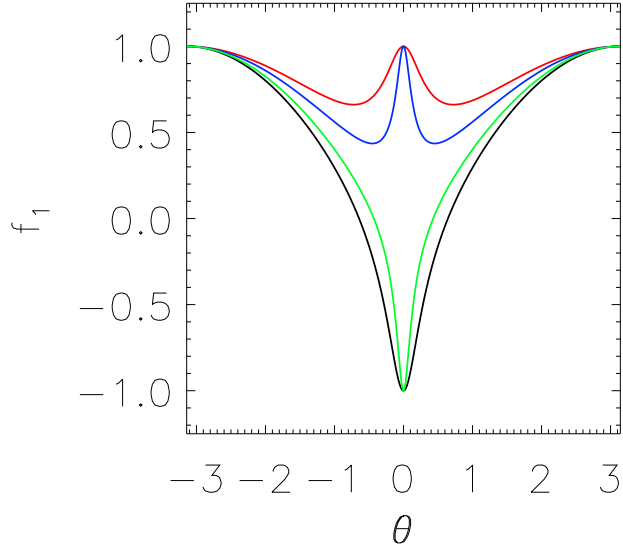,width=10cm}}
\caption{The variations of the field $f1$ for different position
of the substructure. In all plots the mass of the substructure is normalized to $m_p=1$  and the substructure is supposed to be on the X-axis ($\theta_p=0$), 
and at the following distances from the center of the coordinate system: $r_p=1.25$ (black), $r_p=1.1$ (green), $r_p=0.9$ (blue), $r_p=0.75$ (red).}
\label{plot2}
\end{figure}

 For simplicity, a spherical isothermal model will be adopted for the substructure potential.
 The perturbative approach requires the estimation of the deflection field
 on the critical circle. As in Alard (2007) we re-scale the coordinate system so that the 
 Einstein radius of the un-perturbed lens is situated at  $r=1$. The substructure's
 parameters are the following: its mass within the unit circle, $m_p$, its position 
 in polar coordinates: radius, $r_p=1+dr$, and position angle $\theta_p$, (see Fig. ~\ref{plot1}).
 The perturbation induced by this model of substructure 
 is derived in Eq. ~\ref{f0_f1}. The general
 behavior of the function's $\frac{d f_0}{d \theta}$ and $f_1$ is presented
 in Fig.'s ~\ref{plot2} and ~\ref{plot3} respectively.
\begin{equation} \label{f0_f1}
\begin{cases}
 f_1=\frac{m_p \left(1-r_p \cos \left(\theta-\theta_p \right) \right)}{\sqrt{1-2 r_p \cos \left( \theta -\theta_p \right)+r_p^2}} \\
 \frac{d f_0}{d \theta} = \frac{m_p \left(r_p \sin \left(\theta-\theta_p \right) \right)}{\sqrt{1-2 r_p \cos \left( \theta -\theta_p \right)+r_p^2}} \\
\end{cases}
\end{equation}
 \subsection{Properties of the function's $f_1$ and $\frac{d f_0}{d \theta}$}
\begin{figure}
\centering{\epsfig{figure=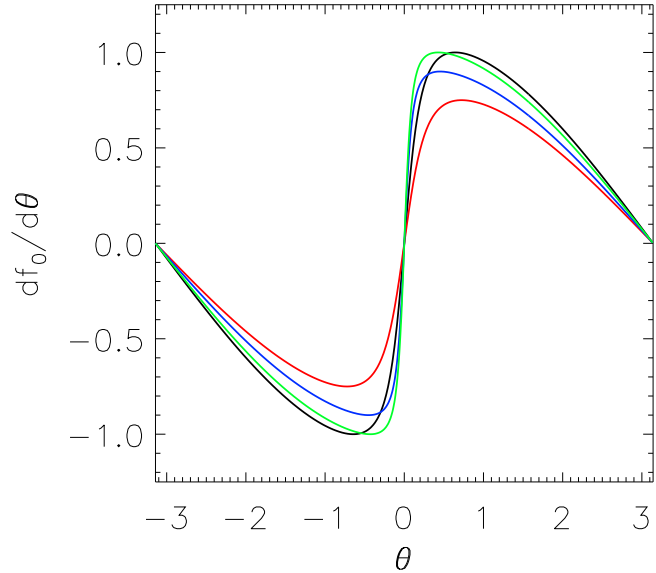,width=10cm}}
\caption{The variations of the field $\frac{d f_0}{d \theta}$, the parameters and color conventions are identical to Fig. ~\ref{plot2} }
\label{plot3}
\end{figure}

 For both function's, the amplitude of the variation is quite similar for different positions of the
 perturbator, and is of the order of the mass of the perturbator. The $f_1$ function is asymmetrical with respect
 to the sign of the $dr$ parameter ($r_p=1+dr$), while for $\frac{d f_0}{d \theta}$ asymmetries are
 weak. The other properties of the function's are related to their steep behavior at the origin. 
 In particular, Fig. ~\ref{plot3} shows that $\frac{d f_0}{d \theta}$, has a steep slope at the origin. This slope $S_0$
 increases as $dr$ decrease, with approximately, $S_0 \simeq 1+\frac{1}{dr}$, which corresponds to a typical scale
 length $\simeq 1-\frac{1}{dr}$. The $f_1$ functional scale like
  $\frac{d f_0}{d \theta}$ near the origin, at lowest order $f_1$ is quadratic and the coefficient of the
 quadratic term is: $\left(1+\frac{1}{dr}\right)^2$, thus the typical scale
 length is also:  $\simeq 1-\frac{1}{dr}$.
 \subsubsection{Spectral decomposition of the function's.}
\begin{figure}
\centering{\epsfig{figure=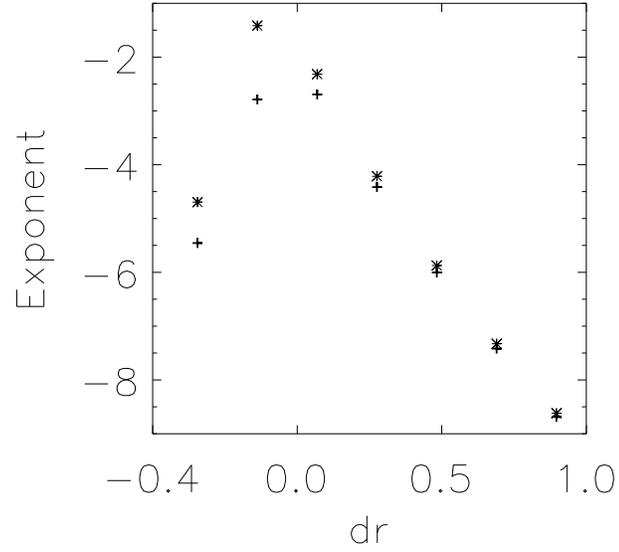,width=10cm}}
\caption{Power law approximation of the power spectrum of $f_1$  and $\frac{d f_0}{d \theta}$ . The Exponent
of the power law is plotted as a function of the distance of the substructure to the unit circle.}
\label{plot4}
\end{figure}

\begin{figure}
\centering{\epsfig{figure=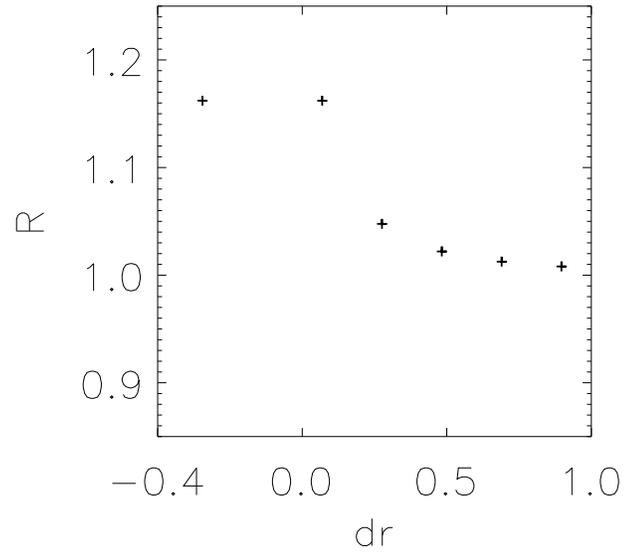,width=10cm}}
\caption{This figure presents
 the ratio of the power spectrum's,  $R = \frac{P_1(n)}{P_0(n)}$.}
\label{plot4.1}
\end{figure}

 The function's   $f_1$ and $\frac{d f_0}{d \theta}$ are expanded in discrete Fourier series, and the power
 spectrum $P_i(n)$ is derived from the coefficients of the Fourier expansion:
\begin{equation}
\begin{cases}
 \frac{d f_0}{d \theta} = \sum_{n} \alpha_{0,n} \cos(n \theta) + \beta_{0,n} \sin(n \theta)\\
 f_1= \sum_{n} \alpha_{1,n} \cos(n \theta) + \beta_{1,n} \sin(n \theta) \\
 P_i(n) = \alpha_{i,n}^2+\beta_{i,n}^2  \ \ \ \  i=0,1 \\
 \label{pot_def}
\end{cases}
\end{equation}
 The power spectrum of the two function's is well approximated by a power law. The exponent of this power-law
 depends strongly on the minimum distance of the sub structure to the Einstein circle (see Fig. ~\ref{plot4}).
 The power-law exponent in itself is variable as a function of $dr$, but interestingly the ratio of the power
 law components is quite constant and close to unity for a wide range of $dr$, see  Fig. ~\ref{plot4.1}.											
\subsection{Relation to the multipole expansion of the potential}
The two fields $f_1$ and $\frac{d f_0}{d \theta}$ are related to the multipole expansion of the 
perturbative potential $\psi$ (see Eq. 3 in Alard 2007 for the definition of $\psi$).
This relation is interesting, since some of the properties of multipole expansion may be exploited in
the analysis of the perturbative function's. 
The expansion of the perturbative potential $\psi$ reads (Kochanek 1991):
\begin{equation} \label{harm_exp}
 \psi = -\sum_n \left( \frac{a_n(r)}{r^n} \cos n \theta + \frac{b_n(r)}{r^n} \sin n \theta + c_n(r) \ r^n \cos n \theta + d_n(r) \ r^n \sin n \theta \right)
\end{equation}
The coefficients $(a_n,b_n,c_n,d_n)$ are related to the density of the lens $\rho$ by the following formula:
\begin{equation} \label{harm_coeffs}
\begin{cases}
a_n = \frac{1}{2 \pi n} \int_0^{2 \pi} \int_0^{r=1} \rho(u,v) \cos n v  \ u^{n+1} \ du \ dv \\
b_n = \frac{1}{2 \pi n} \int_0^{2 \pi}  \int_0^{r=1} \rho(u,v) \sin n v  \ u^{n+1} \ du \ dv \\
c_n = \frac{1}{2 \pi n} \int_0^{2 \pi} \int_{r=1}^\infty \rho(u,v) \cos n v  \ u^{1-n} \  du \ dv \\
d_n = \frac{1}{2 \pi n} \int_0^{2 \pi} \int_{r=1}^\infty \rho(u,v) \sin n v \ u^{1-n} \ du \ dv \\
\end{cases}
\end{equation}
Using Eq's (~\ref{harm_exp}) and  (~\ref{harm_coeffs}), and noting that: $\left(\frac{d \left( a_n+c_n \right)}{d r}\right)_{[r=1]}=\left(\frac{d \left( b_n+d_n \right)}{d r}\right)_{[r=1]}=0$ the fields $f_0$ and $f_1$ :
 \begin{equation} \label{f1f0_harm}
  \begin{cases}
  f_1 = \left(\frac{\partial \psi}{\partial r}\right)_{[r=1]} = \sum_n n \left(a_n-c_n \right) \cos n \theta 
  + n \left(b_n-d_n \right) \sin n \theta\\
  \frac{d f_0}{d \theta} = \left(\frac{\partial \psi}{\partial \theta}\right)_{[r=1]} =  \sum_n -n \left(b_n+d_n \right) \cos n \theta 
  + n \left(a_n+c_n \right) \sin n \theta \\
 \end{cases}
 \end{equation}
\subsubsection{Local perturbation.}
Eq. ~\ref{f1f0_harm} shows that the multipole expansion of the potential is directly related to the harmonic
expansion of the perturbative fields. A simple and interesting case is the perturbation of the potential by a point
mass. There are two cases, either the point mass is inside the unit circle (then form Eq. ~\ref{harm_coeffs}, $c_n=0$ and $d_n=0$) 
or outside ($a_n=0$ and $b_n=0$), in both cases the power spectrum of $f_1$ $P_1(n)$ is equal to the power
spectrum of $\frac{d f_0}{d \theta}$. A substructure is a local perturbation, and is not too far from the point
mass perturbator, this analogy explains the ratio close to 1 which is observed in the ratio between the component of
the power spectrum of the fields (Fig. ~\ref{plot4.1}).
\subsubsection{Slight isophotal deformation of the density.}
Let's consider the case of a lens with small deviations from circular
symmetry. In this case, the total density $\rho_0$ reads:
$$
 \rho_0(r,\theta)=F\left(r \left( 1+g(\theta) \right) \right) \simeq F(r)+r F^{'}(r) \ g(\theta)
$$
With:  $g(\theta) \ll 1$ \\
The perturbative density $\rho$ reads:
\begin{equation} \label{rho_p}
 \rho(r,\theta)=r F^{'}(r) \ g(\theta) 
\end{equation}
 By introducing the former equation in Eq. (~\ref{harm_coeffs}) and subsequently in Eq. (~\ref{f1f0_harm}), 
 the following equations for the fields $f_1$ and $\frac{d f_0}{d \theta}$ are obtained:
\begin{equation}
 \begin{cases}
  f_1 = \sum_n \left(p_n-q_n \right)  \left(\alpha_n \cos(n \theta) + \beta_n \sin(n \theta) \right) \\
  \frac{d f_0}{d \theta} =  \sum_n \left( p_n+q_n \right)  \left(-\beta_n \cos(n \theta) + \alpha_n \sin(n \theta) \right) \\
  
  p_n=\int_0^1 F^{'}(u) u^{n+2} du \\ 
  q_n=\int_1^\infty  F^{'}(u) u^{2-n} du \\ 
  \alpha_n=\frac{1}{2 \pi} \int_0^{2 \pi} g(v) \cos(nv) dv \\
  \beta_n=\frac{1}{2 \pi} \int_0^{2 \pi} g(v) \sin(nv) dv \\
 \end{cases}
\end{equation}
The ratio of the components of the power spectrum of $f_1$, $P_1(n)$, and, $\frac{d f_0}{d \theta}$,  $P_0(n)$ is:
\begin{equation}
 R = \frac{P_1(n)}{P_0(n)} = \left( \frac{p_n-q_n}{p_n+q_n}  \right)^2
\end{equation}
For power law density profiles, $F(r) \propto r^{-\gamma}$, and $R \propto \left( \frac{\gamma-2}{n} \right)^2$, which is very
different from the nearly constant ratio observed for local perturbations.
\section{Properties of the perturbed images.}
\section{Properties of the images formed by perturbed lenses.}
\subsection{Caustics.}
\begin{figure}
\centering{\epsfig{figure=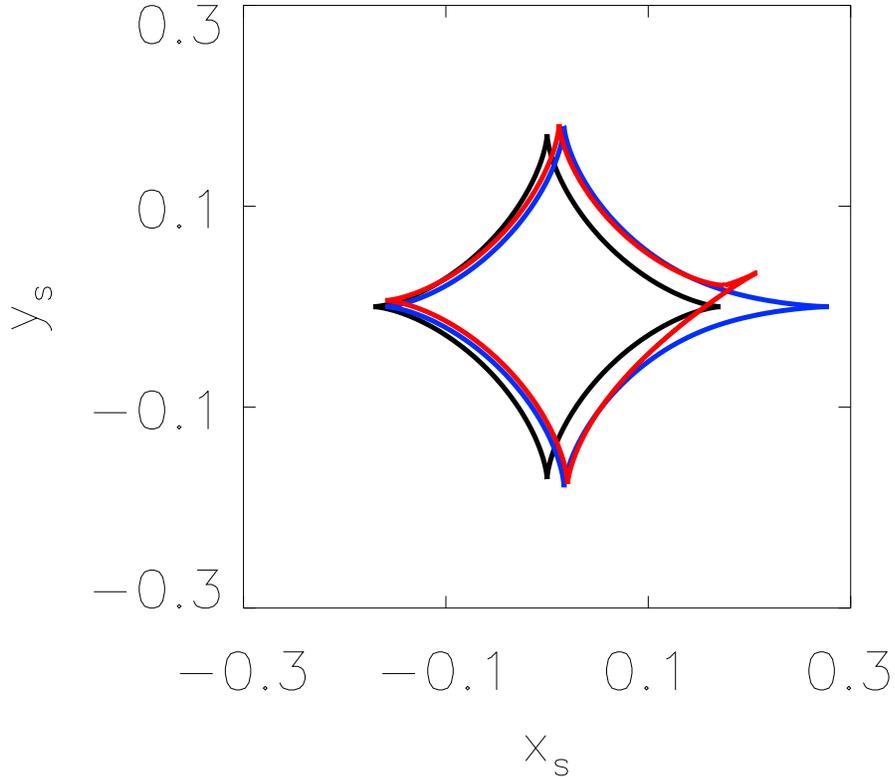,width=14cm}}
\caption{Caustic lines, elliptical lens, $\eta=0.1$ (black), and substructure with parameters, $m_p=0.03$, $\theta_p=0$, $r_p=1.3$ (blue), and substructure,
  $m_p=0.03$,  $theta_p=\frac{\pi}{10}$, $r_p=1.3$ (red). The distortion of the caustic system is maximal when the
 the position of the substructure is aligned with the axis of the elliptical potential.} 
\label{plot5}
\end{figure}

This section will investigate the caustic system of an elliptical lens perturbed by a substructure.
 The lens ellipticity is defined
 by the parameter $\eta$, the substructure parameters are its position angle $\theta_p$, its
distance $dr$  (see Fig. ~\ref{plot1}) and its mass $m_p$ (see Eq. ~\ref{f0_f1}).
 For small $\eta$, the elliptical potential reads:
$$
 \phi_E=F\left(\sqrt{(1-\eta)  \ x^2+(1+\eta) \ y^2} \right) \simeq F(r)-\frac{\eta}{2} \ F^{'}(r) \ r  \cos(2 \theta)=\phi_0(r)+\psi_E(r,\theta)
$$
Adding the contribution of the substructure  (Eq. ~\ref{f0_f1}), and considering that $F^{'}(1)=1$, and $F^{''}(1)=0$ if the model
is isothermal, the perturbative fields are:
\begin{equation}
 \begin{cases} \label{full_pot}
  f_1=-\frac{\eta}{2} \cos(2 \theta)+\frac{m_p \left(1-r_p \cos \left(\theta-\theta_p \right) \right)}{\sqrt{1-2 r_p \cos \left( \theta -\theta_p \right)+r_p^2}} \\
 \frac{d f_0}{d \theta} = \eta \sin(2 \theta)+\frac{m_p \left(r_p \sin \left(\theta-\theta_p \right) \right)}{\sqrt{1-2 r_p \cos \left( \theta -\theta_p \right)+r_p^2}} \\
 \end{cases}
\end{equation}
By introducing the former analytical model in Eq. (31) from Alard (2007) we obtain analytical equations for the caustic lines.
The resulting caustic lines are presented in Fig (~\ref{plot5}). The effect of the substructure is maximal when the substructure
is aligned with the potential axis, and increases with decreasing $dr$. The effect on the caustic is only weakly dependent on the
sign of $dr$. 
\subsection{An illustration of image anomaly due to sub-structure: sub-critical regime.}
The former section shows that the caustic structure is significantly modified in the vicinity of the substructure.
An image configuration situated near the critical line for a purely elliptical lens is shifted to the sub-critical
regime by the substructure field. An interesting point is that this sub-critical regime
is quite different from the regime observed for the purely elliptical lens. The main feature of the sub-critical
regime for perturbed elliptical lens is that the image situated at shorter distance from the substructure is
much more perturbed than the others. Thus by comparing the structure of the images it is generally possible to
detect the anomaly induced by the perturbator. The most obvious features of the perturbed image will be related
to the properties of the function's $f_1$ and $\frac{d f_0}{d \theta}$ (see Fig.'s ~\ref{plot2} and ~\ref{plot3}).
In particular the function's presents large values of the slope ( $\frac{d f_0}{d \theta}$), or large curvature
 ($f_1$) near the origin. For smaller images the local slope is directly related to the image size, thus the perturbation
will reduce the size of the image. For instance let's consider the following sub-critical configurations,
 (i): a source at mid-distance from the caustic in an elliptical potential, ($x_S=x_0$, $y_S=0$), and (ii): the same system, but with the additional
deflection field of a substructure with position angle aligned with the potential axis. In case (i) the sub-critical regime
breaks the arcs in 3 small images, the central image ($\theta=0$), and 2 symmetrical images. For circular sources and a local linear
approximation of the field the size of the images is directly proportional to the inverse of the local 
derivative of  $\frac{d f_0}{d \theta}$. Once the impact parameter is taken into account, the effective
field in the elliptical case is: $\frac{d f_0}{d \theta}=\eta \sin(2 \theta)-x_0 \sin(\theta)$ (see Alard 2007, Eq. 10).
Using the former formula for  $\frac{d f_0}{d \theta}$ some simple calculations shows that the ratio of the size of the
central $S_0$ image to the size of the other images $S_1$ is:
\begin{equation} \label{size_ratio}
 S_{01}=\frac{S_0}{S_1}= 1+\frac{x_0}{2 \eta}
\end{equation}
Eq.  (~\ref{size_ratio}) shows that for sub-critical regime, $0<x_0<2 \eta$, thus $S_{01}>1$ consequently, the central image
is larger that the two other symmetrical images, which is an important feature of the sub-critical regime for elliptical
lens.
With a substructure, the slope near the origin of $\frac{d f_0}{d \theta}$ is perturbed by an additional
field with slope at origin $m_p \left(1+\frac{1}{dr} \right)$ (see Sec. 2.2). In the hypothesis of a local perturbation
the 2 other images remain unperturbed. Assuming $dr \ll 1$ the size of the central image is now:
\begin{equation} \label{size_ratio_pert}
 S_0^{-1} \propto \frac{m_p}{dr} + \left(2 \eta-x_0 \right)
\end{equation}
The scale of $x_0$ in the caustic system is $\eta$, thus Eq.'s  (~\ref{size_ratio_pert}) indicates that
the image ratio is modified by:
\begin{equation}  \label{size_ratio_pert2}
 S_{01}^{P}=S_{01} \times \frac{\eta}{\eta+\frac{m_p}{dr}}
\end{equation}
Thus the perturbation is significant if at least, $m_p \simeq \eta \ dr$. Since the usual scale of both $\eta$ and $dr$
is of about one tenth of $R_E$, $m_p$ must be of only the order of a percent to alter very significantly the size of the
central image. 
\subsection{Analysis of images by Reconstruction of the perturbative field.}  
This section will show that the image anomalies due to substructures can be reconstructed
independantly of the source shape. The reconstruction will be illustrated by the configuration
presented in Sec. 4.1. In this configuration,
the distortion is very obvious, because most of the effect is on the central image, and that
the size of this image is reduced by the field of the substructure. 
However this results holds for source with circular
 contours only. For other
sources the size of the images may be modified slightly, and for better accuracy a more general approach is required.
Here we have to tackle the general problem of lensing potential and source reconstruction, taken together these
problem are difficult and may be quite degenerate. The problem in itself is much simplified if we make the hypothesis
that the field is smooth at the scale of the image and can be represented by a lower order polynomial in $\theta$.
The knowledge of the field at the images positions allows to transfer the image contours to the source plane,
where the different images of the sources must be identical. The constraint that the image must be identical
in the source plane will be used to evaluate the field at the image position. Another constraint on lens reconstruction
is that no images are produced in void areas (Diego {\it et al.}). This constraint is simple to implement
in the perturbative approach, it is sufficient that $|\frac{d f_0}{d \theta}| > R_C$, where $R_C$ is the radius of a circular
envelope to the source contour. 
\subsubsection{Local field models.}
There are basically two kind of image models to consider, smaller images, with nearly linear field at the scale of the
image, and the longer images produced in near caustics configurations . For smaller images, the model to adopt for the $f_0$ field
is very obvious, $\frac{d f_0}{d \theta} \simeq \alpha_0 (\theta -\beta_0)$, with $\alpha_0$ a local constant to evaluate, and 
$\beta_0$ a position angle which in practice should be close to the image center. Near caustics the model should be
a polynomial of higher order. For the $f_1$
field the situation is similar to the $f_0$ field for small images, while in near caustics situations the local modelisation
of the $f_1$ field will require a full polynomial expansion to higher order (2 or 3).
\subsubsection{Fitting local field models.}
\begin{figure}
\centering{\epsfig{figure=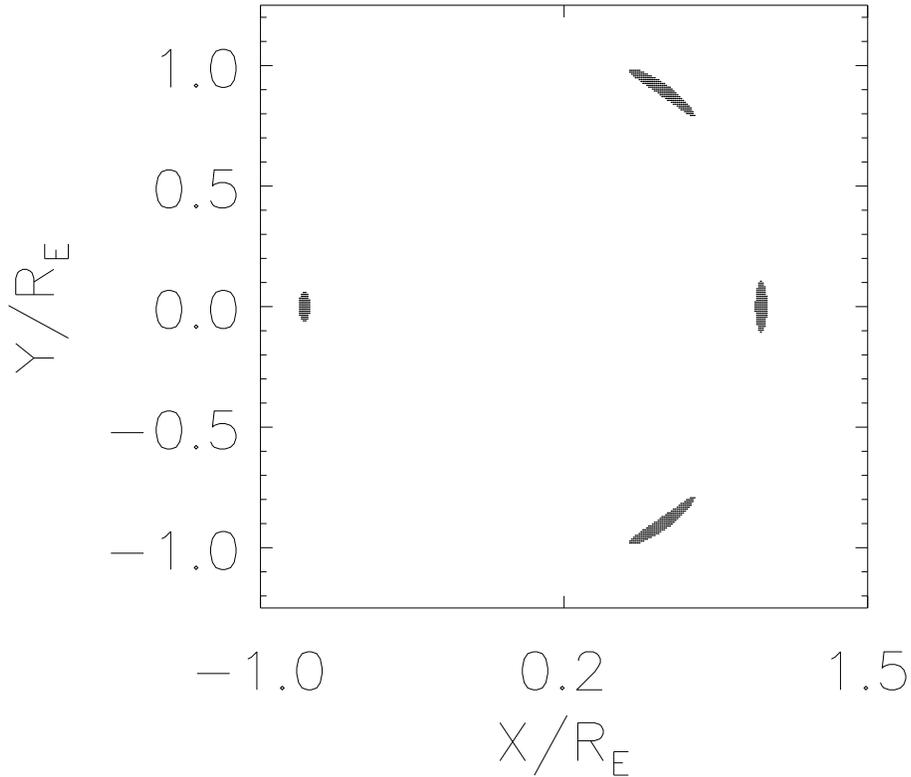,width=14cm}}
\caption{Image obtained by ray-tracing for a circular source. The lens is elliptical, $\eta_S=0.1$, and is perturbed by a substructure with
parameters: $m_p=0.03$, $theta_p=0$, $r_p=1.3$. The source has diameter $R_S=0.05 R_E$, and impact parameter, ($X_S=0.15 R_E$,0). }
\label{plot6}
\end{figure}

\begin{figure}
\centering{\epsfig{figure=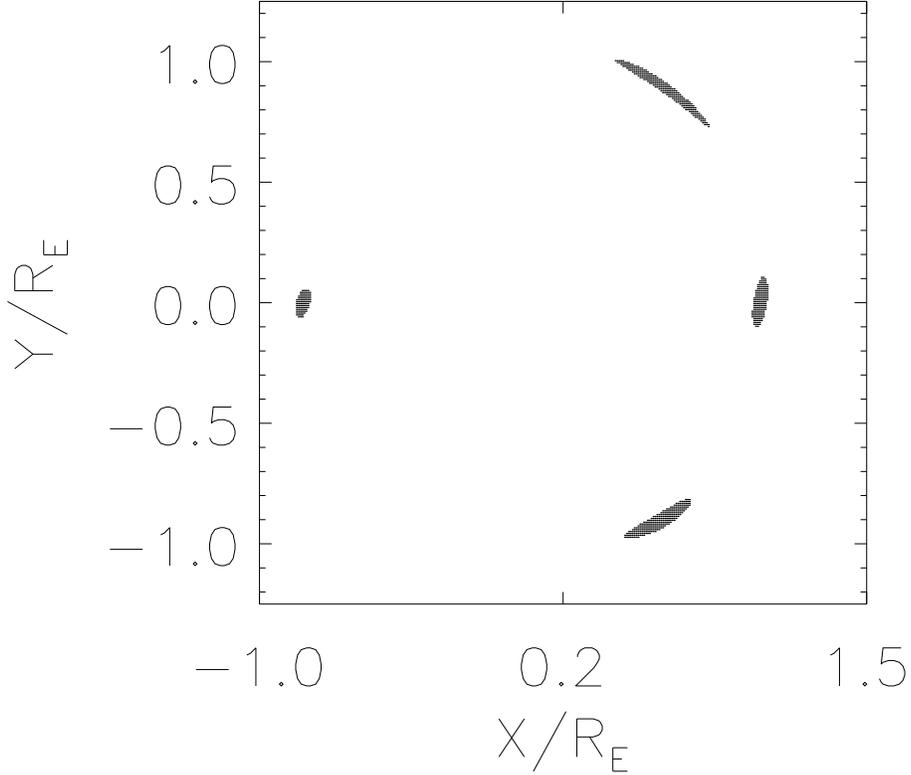,width=14cm}}
\caption{Same as figure ~\ref{plot6}, but for an elliptical source with ellipticity $\eta_S=0.5$}
\label{plot7}
\end{figure}

\begin{figure}
\centering{\epsfig{figure=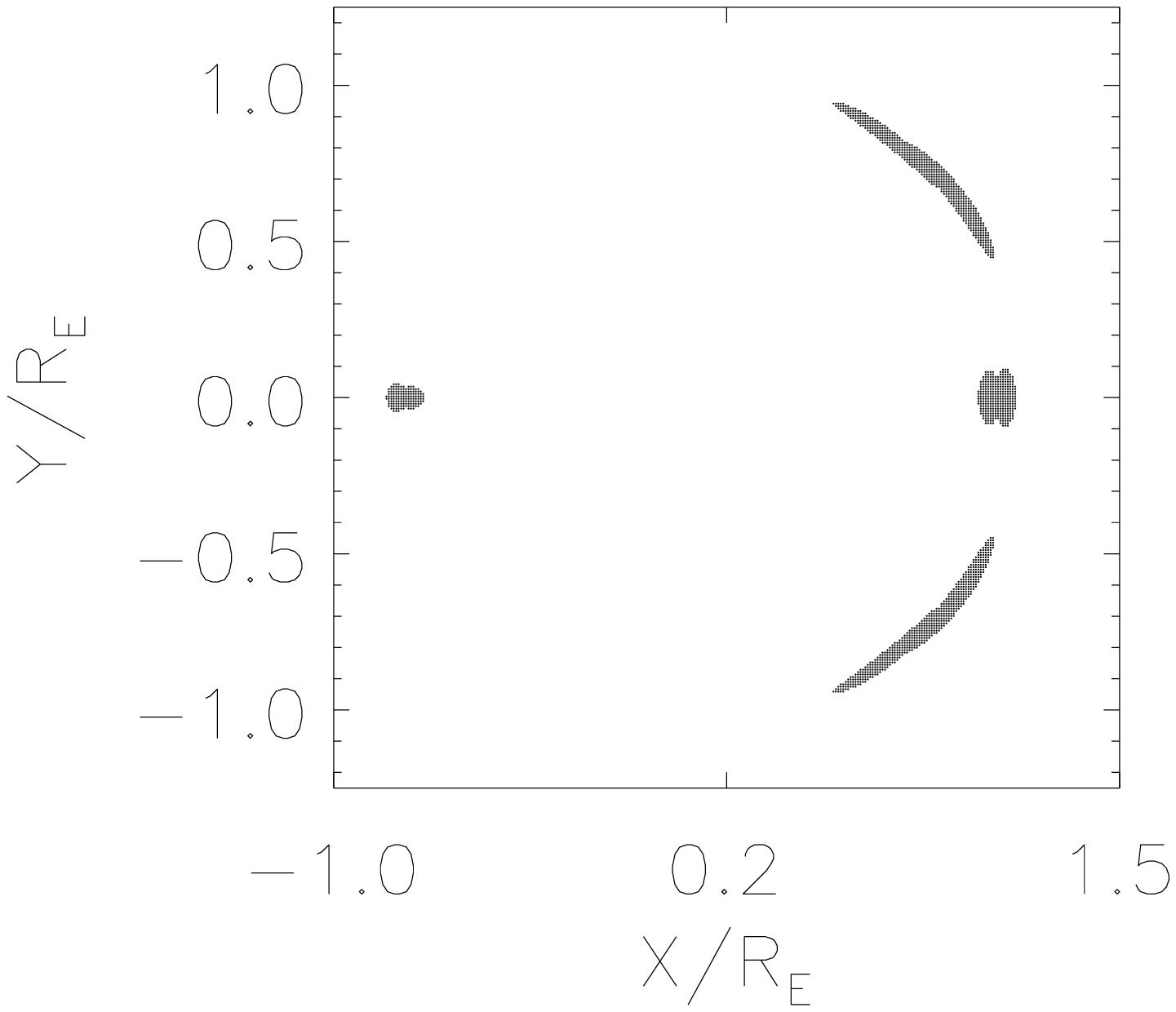,width=14cm}}
\caption{Same as figure ~\ref{plot6}, but for two circular sources with diameter $R_S=0.05$ and respective centers: ($-\frac{R_S}{2}$,0),  ($\frac{R_S}{2}$,0).}
\label{plot8}
\end{figure}

Simulated data are obtained by ray-tracing the lens model in the sub-critical configuration described in Sec. 4.2.
The ellipticity of the lens is $\eta=0.1$, and the substructure parameters are:, $m_P=0.03$, $r_P=1.3$, $\theta_P=0$.
The source position is $x_S=1.5 \eta$, $y_S=0$.
The fitting procedure is greatly simplified by the fact that in the perturbative approach the solution for a circular
source is known (Alard 2007). For circular source the $f_1$ field corresponds to the mean radial position of the contour
at a given $\theta$, and the $\frac{d f_0}{d \theta}$ field is the contour width. Once these fields are extracted
from the images data, they make a good starting guess for the real solution. Basically the guess will be estimated
by fitting local polynomial expansions of $\theta$ to the circular solution. Starting from this guess a few
steps of non-linear fitting optimization will lead to the correct solution. The quantity to minimize to achieve
an estimate of the solution will be a measurement of the image similarity when a re-mapping to the source
plane is performed. A general method to estimate the image similarity is to compare their moments. Suppose that there are $N_I$ images,
and that their moments to order $N_S$ in the source plane are identical, this provide $\frac{N_S (N_S+3) (N_I-1)}{2}$ constraints. 
Local linear image models have $2 N_I$ parameters, thus even in the case of 2 images only the problem is already over-constrained for $N_S=2$.
In the case presented in Fig's (~\ref{plot6}),  (~\ref{plot7}) and  (~\ref{plot8}), there are 4 images, thus moments up to
the second order are sufficient to fit the data. Ray tracing of the sub-critical configuration presented in Section 4.2 is performed
for different sources shape, resulting in 3 different sets of 4 images. Local linear field models are fitted to each of these
sets. The circular solution is used as a first guess and a quantity that measures the distance between the image moments in the
source plane is minimized using the simplex method. In each case both the field parameters and source shape can be recovered (See table ~\ref{res}). 
The great advantage of the local field method, is that even for the higher order fields generated by a substructure, the local
models are simple and can be recovered. Note also that for instance the slope of $\frac{d f_0}{d \theta}$ gives 4 constraints,
which coupled with the image positions gives 8 constraints. In the case of an elliptical lens the Fourier expansion is of order 2
(provided ellipticity is not too large), which corresponds to 4 parameters only. Thus the elliptical case is over-constrained in the
present situation. With only 2 images the elliptical lens is already constrained. Taking any 2 images unperturbed by the substructure, and making
an elliptical model, it will always be impossible for this model to meet the constraint related to the perturbed image. 
\begin{table}[ht]
\caption{Local model fitting results for different sources, first circular source with diameter $R_S=0.05 R_E$, elliptical
source with $\eta_S=0.5$, $(1-\eta_S) x_S^2+(1-\eta_S) y_S^2=R_S^2$, and $R_S=0.05 R_E$, two circular sources with diameter $R_S=0.05 R_E$ and respective centers: ($-\frac{R_S}{2}$,0),  ($\frac{R_S}{2}$,0). For comparison, 
in the last column
of the first part of the table, the theoretical results of the local field values are given
 when no substructure is present (see the large difference on the slope of image 2). It is not presented
for the second part, since due to symmetry of the substructure position it has no effect on the $f_1$ field. }
\centering
\begin{tabular}{c c c c c}
$|\frac{d f_0}{d \theta}|$ & Slope image 1 &   Slope image 2  & Slope image 3  & Slope image 4 \\
\hline
True solution & 0.17 &  0.23 & 0.17 & 0.35 \\
 circular source & 0.17 & 0.25 & 0.17 & 0.34 \\
 Elliptical source & 0.2 & 0.26 & 0.19 & 0.37 \\
2 circular sources & 0.17 & 0.28 & 0.16 & 0.38 \\
Without substructure & 0.1 & 0.06 & 0.1 & 0.36 \\
\hline
\hline
$f_1$ & Slope image 1 &   Slope image 2  & Slope image 3  & Slope image 4 \\
\hline
True solution & 0 &  0 & 0 & 0 \\
circular source & 0.01 & 0 & -0.01 & 0 \\
Elliptical source & 0.04 & -0.05 & -0.05 & 0.03 \\
2 circular sources & 0.02 & -0.03 & -0.01 & -0.05 \\
\hline
\end{tabular}
\label{res}
\end{table}
\subsubsection{Accuracy of measurements using the perturbative method.}
In Table 1 the perturbation on the local slope of the $\frac{d f_0}{d \theta}$ function due to the substructure is about 7 times
the mean scatter of the measurements recovered by fitting local models for the different sources. Translated in accuracy on the measurement
of the substructure mass gives about 0.4 \% of the main halo mass. However the scatter in the measurements comes from the perturbative approximation and also from
the simplicity of the local (linear) modeling. Thus this scatter is an over-estimate of the error made in the perturbative 
approximation.
The perturbative method may introduce some limitation in accuracy, but in practice, resolution effects and noise should be much 
stronger limiting factors. And more importantly, the limit in accuracy by the perturbative method may be a concern only for
absolute measurement of the displacement field of the lens, if we are interested in the differential effect of the substructure
field, then the perturbative method will be very accurate, first because the method is linear, and thus allows differential measurements,
 and second because also due to linearity, the field of the perturbator can be reconstructed with an accuracy
 that scales likes its mass. Thus, for the evaluation of the differential effects due to very small substructure the perturbative
method should give accurate results, in this case, the main problem will be related to the un-biased statistical estimation of the perturbative fields
 for distributions without substructure.
\subsection{Approximate source invariant quantities}
In general the image features are dependent upon the source shape, and the re-construction of the lens fields require
the non-linear procedure presented in the former section. However, for smaller images and weakly elliptical sources, there
is an approximate invariant. This conserved quantity is useful for nearly round sources, or for improving the first
guess in the non linear fitting procedure. Let's consider small images with total size $\theta_I \ll 1$ and an elliptical source,
the width of the image is given by Eq. (15) in Alard (2007):
\begin{equation}  \label{width}
\begin{cases}
W=\frac{\sqrt{R_0^2 \ S-(1-\eta_S^2) \ \left( \frac{d f0}{d \theta} \right)^2}}{S} \\
S=1-\eta_S \cos\left(2 \left(\theta-\theta_0 \right) \right)
\end{cases}
\end{equation}
Where $\eta_S$ is the source ellipticity and $\theta_0$ is the angle of the source main axis. Since the image
is supposed to be small, we are operating in a small range of $\theta$, and the field $ \frac{d f0}{d \theta}$
may be linearized locally. Near the center of the image,  $ \frac{d f0}{d \theta} \simeq 0$, thus by taking
the origin of $\theta$ at the image center we have:  $ \frac{d f0}{d \theta} \simeq k \theta$. We will also
assume that the ellipticity is a small number, so that by change of variable $\eta_S = \epsilon \eta_S$, and
$\theta=\epsilon \theta$. Using these new variables, it is possible to expand Eq. (~\ref{width}) in series of $\epsilon$, which
simplifies the calculation of many quantities.
. In particular, the image size along the orthoradial direction is obtained by the condition $W=0$.
Solving to the lowest order in $\epsilon$, we obtain a second order equation in $\theta$, and
the difference of the 2 roots gives
the image size. To the lowest order in $\epsilon$, the image size in the orthoradial direction is $W_T$:
\begin{equation}
 W_T=\frac{R_0}{k} \left(1-\frac{ \eta_S \cos 2 \theta_0}{2} \right)
\end{equation}
The size of the image in the other direction is approximately the size of the image in the radial
direction near the center of the image, from Eq. (~/ref{width}) to the lowest order in $\epsilon$,
the radial size $W_R$ is:
\begin{equation}
 W_R=\frac{R_0}{k} \left(1+\frac{ \eta_S \cos 2 \theta_0}{2} \right)
\end{equation}
To first order in $\eta_S$, the product $S=W_T \times W_R$ which is closely related to the image surface 
does not depend on the ellipticity of the source.
This result means that in practice for small ellipticity ($\eta_S \ll 1$), $S$ is a constant independent of
the source ellipticity.


\begin{thebibliography}{}
%
\bibitem[]{} Alard, C., 2007, MNRAS Letters, 382, 58
\bibitem[]{} Bartelmann, Matthias, Steinmetz, Matthias, Weiss, Achim, 1995, A\&A, 297, 1 
\bibitem[BK1987]{} Bartelmann, M., 1996, A\&A, 313, 697 
\bibitem[]{} Diego, J.M.,Protopapas, P., Sandvik, H.B., Tegmark, M., 2005, MNRAS, 360, 477
\bibitem[]{} Ghigna, S., Moore, B., Governato, F., Lake, G., Quinn, T., Stadel, J., 2000, ApJ, 544, 616
\bibitem[]{} Horesh, A., Ofek, E.O., Maoz, D., Bartelmann, M., Meneghetti, M., Rix, HW., 2005, ApJ, 633, 768
\bibitem[]{} Keeton, C.R., Gaudi, B.S., Petters, A.O., 2003, ApJ, 598, 138
\bibitem[]{} Klypin, A., Kravtsov, A, V., Valenzuela, O., Prada, F., 1999, ApJ, 522, 82
\bibitem[BK1991]{} Kochanek, C. S., 1991, ApJ, 373, 354 
\bibitem[]{} Maccio, A.V., Moore, B., Stadel, J., Diemand, J., 2006, MNRAS, 366, 1529
\bibitem[]{} Meneghetti, M., Argazzi, R., Pace, F., Moscardini, L., Dolag, K., Bartelmann, M., Li, G., Oguri, M., 2007, A\&A, 461, 25 
\bibitem[]{} Moore, B. Ghigna, S., Governato, F., Lake, G., Quinn, T., Stadel, J., Tozzi, P., 1999, ApJL, 524, 19
\bibitem[]{} Mao, S., J., Y., Ostriker, J.P. Weller, J., 2004, ApJL, 604, 5
\bibitem[]{} Peirani, S., Alard, C., Pichon, C., Gavazzi, R., Aubert, D., 2008, in preparation.
%
\end{thebibliography}
\end{document}